\documentclass[12pt]{iopart}
\pdfoutput=1
\usepackage{graphicx}

\begin{document}

\title[Accidental line nodes in two-dimensional materials]{Movable but unavoidable nodal lines through high-symmetry points in two-dimensional materials}

\author{V Damljanovi\'c}

\address{Institute of Physics Belgrade, University of Belgrade, Pregrevica 118, 11080 Belgrade, Serbia}
\ead{damlja@ipb.ac.rs}
\vspace{10pt}
\begin{indented}
\item[]February 2023
\end{indented}

\begin{abstract}
In two-dimensional materials electronic band contacts often give non-trivial contribution to materials topological properties. Besides band contacts at high-symmetry points (HSP) in the Brillouin zone (BZ), like those in graphene, there are nodal lines which form various patterns in the reciprocal space. In this paper we have found all movable nodal lines, which shape depends on the model, that pass through HSPs in the presence of time-reversal symmetry. Cases with and without spin-orbit coupling are included by studying all eighty layer groups and their double extensions. Eight single and six double groups, including three symmorphic, necessarily host Dirac and Weyl nodal lines that extend through the whole BZ, respectively. Our research might be of interest in designing new materials with interesting physical properties.
\end{abstract}

%
%
%
%
%

\section{Introduction}
Physics of two-dimensional (2D) materials became one of the focal points of solid-state science, since the electric field effect has been reported in atomically thin graphene in 2004. Brillouin zone (BZ) of 2D materials is an even-dimensional manifold, so topological invariants are used in explaining properties of 2D materials - such as integer quantum Hall effect. Non-zero topological charge originates from singularities of electron wave functions in the reciprocal space, and these singularities appear at BZ points where two or more electron bands touch \cite{Rev2D}.

In 2D, bands can touch at isolated points or lines. Band touchings can be essential \cite{BSW36, HerrTRS} or accidental \cite{HerrAccBC}. The former are induced by the crystal and time-reversal symmetry (TRS), they are pinned to a high symmetry point (HSP) or line (HSL) and they are intact by symmetry preserving perturbations. Accidental band contacts change their position in the BZ by perturbations that keep the symmetry, they cannot be removed if protected by topology, or can be removed in the opposite case (truly accidental band contacts). So far, the existence of such movable but unavoidable nodal points (MNP) and lines (MNL) was almost exclusively linked to non-symmorphic symmetries of a crystal \cite{YK15, WiKa16, AccBCh, AccBCt, AccBCo}. We show below that MNLs exist in some symmorphic groups too.

Nodal lines are further classified according to dispersion in the directions perpendicular to the line. For higher order nodal lines, the splitting is of order higher then one. In 3D non magnetic materials with spin-orbit coupling (SOC) such nodal lines are of order two and three \cite{QCNL19}. Weyl (WNL) and Dirac nodal lines (DNL) have linear splitting, with two-fold and four-fold degeneracy on the line, respectively. Existence of WNL or DNL in bulk band structure leads to interesting physical properties, such as \emph{drumhead} edge states, which are nearly dispersionless and hence exhibits strong electron correlations \cite{VeLiR1, VeLiR2}. Existence of nodal lines in 3D is connected with particular symmetry element \cite{NLS1, NLS2, NLS3, NLS4, NLS5, NLS6}, and also analyzed by use of point groups \cite{NLS7}, all 230 space groups \cite{NLS8} and with 3D magnetic groups \cite{NLSmag}. These publications are augmented with exhaustive tables of all possible quasiparticles in 3D magnetic and non-magnetic crystals \cite{Enci1, Enci2, Enci3, Enci4, Enci5}.

In 2D WNLs are predicted in $\mathrm{GaTeI}$ family \cite{NL2D1} and in ferromagnetic layers \cite{NL2Dfm1, NL2Dfm2, NL2Dfm3}. For determination of properties of 2D materials using symmetry, characters or irreducible (co)representations of layer groups are necessary. Those are tabulated for layer double little groups \cite{gdg}, for generators of layer single groups \cite{Dza98} and for maximal symmetry of hexagonal materials \cite{Ja14}. Recently, full tabulation of layer groups, single and double, gray and ordinary is published \cite{DGsajt}. We used \cite{DGsajt} together with tables for (3D) space groups from Bilbao Crystallographic Server \cite{BCS06, BCSDG17} as a starting point of our research.

Here we present all MNLs that pass through HSPs in layer groups with TRS, with and without SOC. We show that these MNLs are unavoidable for crystals belonging to eight layer groups (describing systems without SOC), and six layer double groups (for crystals with SOC). We give dispersions and effective Hamiltonians near mentioned HSPs, and have illustrated our theory with few examples. An interesting curiosity are three symmorphic layer groups that support MNLs when SOC is included. Finally, we compare our results with those published in the literature on related topics.

\section{Method}
For $\mathbf{k}_0$ being a BZ point (HSP), $\mathbf{q}$ a small wave vector, $R$ allowed irreducible (co)representation of little group $G_{\mathbf{k}_0}$ of $\mathbf{k}_0$ and $g=\left\{\widehat{h}\right|\left.\mathbf{\tau}_{\hat{h}}\right\}$ an element (in Seitz notation) of $G_{\mathbf{k}_0}$, the following formula holds \cite{Manjes12}:
\numparts
\label{meto}
\begin{equation}
\label{crsym}
\widehat{H}\left(\mathbf{k}_0+\mathbf{q}\right)=\widehat{R}^{\dag}\left(g\right)\widehat{H}\left(\mathbf{k}_0+\widehat{h}'\mathbf{q}\right)\widehat{R}\left(g\right),
\end{equation}
where $\widehat{h}'$ is the operator reduction of $\widehat{h}$ to 2D BZ. If, in addition, $-\mathbf{k}_0$ belongs to the star of $\mathbf{k}_0$, then TRS (denoted by $\theta$) gives:
\begin{equation}
\label{trs}
\widehat{H}'^*\left(\mathbf{q}\right)=\widehat{R}^{\dag}\left(\theta g_s\right)\widehat{H}'\left(-\widehat{h}_s'\mathbf{q}\right)\widehat{R}\left(\theta g_s\right),
\end{equation}
\endnumparts
where $\widehat{H}'(\mathbf{q})=\widehat{H}(\mathbf{k}_0+\mathbf{q})-\widehat{H}(\mathbf{k}_0)$, $g_s=\left\{\widehat{h}_s\right|\left.\mathbf{\tau}_{\hat{h}_s}\right\}$ and $\widehat{h}_s\mathbf{k}_0$ differs from $-\mathbf{k}_0$ by a reciprocal lattice vector. By differentiating these relations with respect to $\mathbf{q}$ at $\mathbf{q}=0$ once, twice,...\emph{etc.} we get relations between Hamiltonian derivatives, and from there we derive dispersions in the vicinity of HSP. We have considered Hamiltonians up to and including second (third) order for case without (with) SOC and we have isolated groups and HSPs that are not located at HSL and that have line-like solutions for vanishing splitting terms. In all such cases, two-component Hamiltonians contain only one Pauli matrix (in addition to the unit matrix) which we choose to be $\widehat{\sigma}_1$. Since $n$-th order Hamiltonian derivatives transform as $[\Gamma_{2DPV}^{\otimes n}]\otimes R^* \otimes R$, and since in all cases of interest (except for double group $74^D$-BZ center), $\Gamma_{2DPV}^{\otimes 3}=4\Gamma_{2DPV}$ (or $\Gamma_{2DPV}^{\otimes 3}=2\Gamma_1+3\Gamma_{2DPV}$, for $74^D$-BZ center), we could exclude the possibility that some higher order derivatives couple to any other Pauli matrix, than the ones already contained in the linear or quadratic (linear, quadratic or cubic for $74^D$-BZ center) terms in Hamiltonian Taylor expansion, which would cause energy splitting. $\Gamma_{2DPV}$ denotes 2D polar-vector representation that acts in 2D reciprocal space, $\Gamma_1$ is the unit representation, while $[\Gamma_{2DPV}^{\otimes n}]$ is the symmetrised $n$-th power of $\Gamma_{2DPV}$. It turned out that four-component Hamiltonians, that do not correspond to fortune teller (FT) dispersion \cite{Ja17, Mi20}, do not support MNLs. Since gray layer single and double groups have multidimensional little group correps of dimensions two and four, our method exhausts all possibilities.

We now prove that, if MNL exists in the vicinity of a HSP, it must continue further. For $\mathbf{k}_0$ now being touching point of two bands located near HSP, its little group contains only horizontal reflection plane as a nontrivial element. Both bands can be approximated with $E_1=p_1q_1+p_2q_2$ and $E_2=c_1q_1+c_2q_2$ respectively, so that $p_1\neq c_1$ and $p_2\neq c_2$ (at $\mathbf{q}=0$ two bands touch by initial assumption, $\mathbf{k}_0$ is now not a HSP but lies close to it). Now $E_1=E_2$ has solution $q_2=q_1(p_1-c_1)/(c_2-p_2)$ which means that the line continues. We can repeat the same reasoning for the another point close to $\mathbf{k}_0$ and so on. In other words, MNL can not be interrupted.

Finally, guided by the result described in the previous paragraph, we have included groups that host FT states, described in \cite{Ja17} and \cite{Mi20} for non-SOC and SOC cases, respectively, and also experimentally observed \cite{Zdyb}. Among conclusions of previous paragraph is that FT states cannot be the only features at the Fermi level. This holds no matter if Chern number of FT states is integer or not.

\section{Results}
In what follows, $c$'s denote real parameters, while $q_1$ and $q_2$ are projections of $\mathbf{q}$ to orthonormal basis vectors $\mathbf{e}_1$ and $\mathbf{e}_2$, respectively. The orientation of basis vectors is arbitrary since no little group of HSPs in question contains horizontal rotation axis or vertical reflection plane, except for groups hosting FT dispersion. However, depending on the parameters of the model, sometimes it is more convenient to switch to other orthonormal basis $\mathbf{e}_1'$, $\mathbf{e}_2'$, which will be defined below. In these cases projections of $\mathbf{q}$ to this new basis are denoted with $q_1'$ and $q_2'$. Finally $V_{l,m}^{(2)}$ ($V_{l,m,n}^{(3)}$) are second (third) order derivatives of certain components of the Hamiltonian. These are real parameters, symmetric under any permutation of indices, but with no additional symmetry. The superscript $D$ denotes double group, which is the symmetry of the system with SOC.

Groups and HSPs hosting MNL when SOC is neglected are shown in the Table \ref{tbl1}, those with SOC in the Table \ref{tbl2}. For groups $5$, $7$, $36$, $48$, $52$(points $(0,\pm1/2)$, $(\pm1/2,0)$), $4^D$, $5^D$, $35^D$ and $74^D$(points $(0,\pm1/2)$, $(\pm1/2,0)$, $\pm(1/2,1/2)$), the Hamiltonian and dispersion are:
\numparts
\label{hadis1}
\begin{equation}
\label{ha1}
\fl \widehat{H}=\left(E_0+\sum_{l,m=1}^2V_{l,m}^{(2)}q_lq_m\right)\widehat{\sigma}_0+\left(c_1q_1+c_2q_2+\sum_{l,m,n=1}^2V_{l,m,n}^{(3)}q_lq_mq_n\right)\widehat{\sigma}_1,
\end{equation}
\begin{equation}
\label{dis1}
\fl E_{1,2}=E_0+\sum_{l,m=1}^2V_{l,m}^{(2)}q_lq_m\pm|c_1q_1+c_2q_2+\sum_{l,m,n=1}^2V_{l,m,n}^{(3)}q_lq_mq_n|.
\end{equation}
For new basis given by:
\begin{eqnarray}
\mathbf{e}_1'=\frac{c_1\mathbf{e}_1+c_2\mathbf{e}_2}{\sqrt{c_1^2+c_2^2}}, \nonumber \\
                                                                          \\
\mathbf{e}_2'=\frac{-c_2\mathbf{e}_1+c_1\mathbf{e}_2}{\sqrt{c_1^2+c_2^2}}, \nonumber
\end{eqnarray}
the Hamiltonian and dispersion are:
\begin{equation}
\label{ha1dr}
\fl \widehat{H}=\left(E_0+\sum_{l,m=1}^2V_{l,m}'^{(2)}q_l'q_m'\right)\widehat{\sigma}_0+\sqrt{c_1^2+c_2^2}\left(q_1'+c_0q_2'^3\right)\widehat{\sigma}_1,
\end{equation}
\begin{equation}
\label{dis1dr}
\fl E_{1,2}=E_0+\sum_{l,m=1}^2V_{l,m}'^{(2)}q_l'q_m'\pm\sqrt{c_1^2+c_2^2}|q_1'+c_0q_2'^3|,
\end{equation}
with $c_0=(c_1^3V_{222}^{(3)}-3c_1^2c_2V_{122}^{(3)}+3c_1c_2^2V_{112}^{(3)}-c_2^3V_{111}^{(3)})/(c_1^2+c_2^2)^2$. Since $q_1'+c_0q_2'^3=0$ has one solution, one MNL passes through $\mathbf{q}=0$. In (\ref{ha1dr}) and (\ref{dis1dr}) we have neglected energy splitting terms that contain products of $q_1'$ with $q_2'$, in comparison to $q_1'$. Such HSPs were mentioned in \cite{Mi22} as having linearity rank one.
\endnumparts

\begin{table}
\caption{\label{tbl1} Layer single groups hosting unavoidable accidental nodal lines through HSP. The notation for layer and space groups as well as primitive basis vectors are according to \cite{vole} and \cite{vola}, respectively. The coordinates of HSPs in the last column, are in the primitive basis $\left\{\mathbf{b}_1, \mathbf{b}_2\right\}$ of the reciprocal 2D lattice.}
\begin{indented}
\item[]
\begin{tabular}{@{}lllllcc}
\br
\multicolumn{2}{l}{Layer single group} & \multicolumn{3}{l}{Corresponding space group} & Diperiodic plane & $HSP$ \\
\mr
5 & $p \ 1 \ 1 \ b$ & 7 & $P \ 1 \ c \ 1$ & $C_s^2$ & $y=0$ & $(0,\pm\frac{1}{2})$\\
  &                 &   &                 &         &       & $(\pm\frac{1}{2},\pm\frac{1}{2})$ \\
\rule{0pt}{3ex}7 & $p \  1 \ 1 \ 2/b$ & 13 & $P \ 1 \ 2/c \ 1$ & $C_{2h}^4$ & $y=0$ & $(0,\pm\frac{1}{2})$\\
  &                 &       &                 &         &       & $(\pm\frac{1}{2},\pm\frac{1}{2})$ \\
\rule{0pt}{3ex}33& $p \  b \ 2_1 \ a$ & 29 & $P \ c \ a \ 2_1$ & $C_{2v}^5$ & $y=0$ & $(\pm\frac{1}{2},\pm\frac{1}{2})$\\
\rule{0pt}{3ex}36& $c \ m \ 2 \ e$     & 39 & $A \ e \ m \ 2$        & $C_{2v}^{15}$& $x=0$ & $(0,\pm\frac{1}{2})$ \\
  &                 &       &                 &         &       & $(\pm\frac{1}{2},0)$ \\
\rule{0pt}{3ex}43& $p \  2/b \ 2_1/a \ 2/a$ & 54 & $P \ 2_1/c \ 2/c \ 2/a$ & $D_{2h}^8$ & $y=0$ & $(\pm\frac{1}{2},\pm\frac{1}{2})$\\
\rule{0pt}{3ex}45& $p \  2_1/b \ 2_1/m \ 2/a$ & 57 & $P \ 2/b \ 2_1/c \ 2_1/m$ & $D_{2h}^{11}$ & $x=0$ & $(\pm\frac{1}{2},\pm\frac{1}{2})$\\
\rule{0pt}{3ex}48 & $c \ 2/m \ 2/m \ 2/e$ & 67 & $C \ 2/m \ 2/m \ 2/e$ &$D_{2h}^{21}$ & $z=0$ & $(0,\pm\frac{1}{2})$  \\
  &                 &       &                 &         &       & $(\pm\frac{1}{2},0)$ \\
\rule{0pt}{3ex}52 & $p \ 4/n$ & 85 & $P \ 4/n$ & $C_{4h}^3$ & $z=0$ & $(0,\pm\frac{1}{2})$  \\
  &                 &       &                 &         &       & $(\pm\frac{1}{2},0)$ \\
	&                 &   &                 &         &       & $(\pm\frac{1}{2},\pm\frac{1}{2})$ \\
\br
\end{tabular}
\end{indented}
\end{table}

\begin{table}
\caption{\label{tbl2} Layer double groups hosting unavoidable accidental nodal lines through HSP. The notation for layer and space groups as well as primitive basis vectors are according to \cite{vole} and \cite{vola}, respectively. The coordinates of HSPs in the last column, are in the primitive basis $\left\{\mathbf{b}_1, \mathbf{b}_2\right\}$ of the reciprocal 2D lattice.}
\begin{indented}
\item[]
\begin{tabular}{@{}lllllcc}
\br
\multicolumn{2}{l}{Layer double} & \multicolumn{3}{l}{Corresponding space} & Diperiodic & \\
\multicolumn{2}{l}{group} & \multicolumn{3}{l}{double group} & plane & $HSP$ \\
\mr
$4^D$ & $p \ 1 \ 1 \ m$ & $6^D$ & $P \ 1 \ m \ 1$ & $C_s^1$ & $y=0$ & $(0, 0)$\\
      &                 &       &                 &         &       & $(\pm\frac{1}{2},0)$ \\
	    &                 &       &                 &         &       & $(0,\pm\frac{1}{2})$ \\		
      &                 &       &                 &         &       & $(\pm\frac{1}{2},\pm\frac{1}{2})$ \\
\rule{0pt}{3ex}$5^D$ & $p \  1 \ 1 \ b$ & $7^D$ & $P \ 1 \ c \ 1$ & $C_s^2$ & $y=0$ & $(0,0)$\\
      &                 &       &                 &         &       & $(\pm\frac{1}{2},0)$ \\
\rule{0pt}{3ex}$29^D$ & $p \  b \ 2_1 \ m$ & $26^D$ & $P \ m \ c \ 2_1$ & $C_{2v}^2$ & $x=0$ & $(0,\pm\frac{1}{2})$\\
      &                 &       &                 &         &       & $(\pm\frac{1}{2},\pm\frac{1}{2})$ \\
\rule{0pt}{3ex}$33^D$& $p \  b \ 2_1 \ a$ & $29^D$ & $P \ c \ a \ 2_1$ & $C_{2v}^5$ & $y=0$ & $(0,\pm\frac{1}{2})$\\
\rule{0pt}{3ex}$35^D$& $c \ m \ 2 \ m$  & $38^D$ & $A \ m \ m \ 2$    & $C_{2v}^{14}$& $x=0$ & $(0,\pm\frac{1}{2})$ \\
      &                 &       &                 &         &       & $(\pm\frac{1}{2},0)$ \\
\rule{0pt}{3ex}$74^D$& $p \ \overline{6}$& $174^D$ & $P \ \overline{6}$ &$C_{3h}^1$ & $z=0$ & $(0,0)$ \\
      &                 &       &                 &         &       & $(\pm\frac{1}{2},0)$ \\
			&                 &       &                 &         &       & $(0,\pm\frac{1}{2})$ \\
			&                 &       &                 &         &       & $\pm(\frac{1}{2},\frac{1}{2})$ \\
\br
\end{tabular}
\end{indented}
\end{table}

For group $52$ near BZ corners, the following relations hold:
\numparts
\label{hadis2}
\begin{equation}
\label{ha2}
\widehat{H}=\left(E_0+c\mathbf{q}^2\right)\widehat{\sigma}_0+\left(c_1q_1^2+c_2q_1q_2-c_1q_2^2\right)\widehat{\sigma}_1,
\end{equation}
\begin{equation}
\label{dis2}
E_{1,2}=E_0+c\mathbf{q}^2\pm|c_1q_1^2+c_2q_1q_2-c_1q_2^2|.
\end{equation}
In the new basis:
\begin{eqnarray}
\mathbf{e}_1'=\frac{\mathbf{e}_1+\lambda\mathbf{e}_2}{\sqrt{1+\lambda^2}}, \nonumber \\
                                                                              \\
\mathbf{e}_2'=\frac{-\lambda\mathbf{e}_1+\mathbf{e}_2}{\sqrt{1+\lambda^2}}, \nonumber
\end{eqnarray}
where $\lambda=(c_2+\sqrt{c_2^2+4c_1^2})/(2c_1)$, the following holds:
\begin{equation}
\label{ha2dr}
\widehat{H}=\left(E_0+c\mathbf{q}'^2\right)\widehat{\sigma}_0-\sqrt{c_2^2+4c_1^2}q_1'q_2'\widehat{\sigma}_1,
\end{equation}
\begin{equation}
\label{dis2dr}
E_{1,2}=E_0+c\mathbf{q}'^2\pm\sqrt{c_2^2+4c_1^2}|q_1'q_2'|.
\end{equation}
\endnumparts
Since $q_1'q_2'=0$ has two solutions, two MNL intersect at right angle for $\mathbf{q}=0$.

For group $74^D$ near the BZ center ($\mathbf{k}_0=0$):
\numparts
\label{hadis3}
\begin{equation}
\label{ha3}
\widehat{H}=\left(E_0+c\mathbf{q}^2\right)\widehat{\sigma}_0+\left(c_1q_1(q_1^2-3q_2^2)+c_2q_2(q_2^2-3q_1^2)\right)\widehat{\sigma}_1,
\end{equation}
\begin{equation}
\label{dis3}
E_{1,2}=E_0+c\mathbf{q}^2\pm|c_1q_1(q_1^2-3q_2^2)+c_2q_2(q_2^2-3q_1^2)|.
\end{equation}
New basis is given by:
\begin{eqnarray}
\mathbf{e}_1'=\frac{c_2\mathbf{e}_1+c_1B\mathbf{e}_2}{\sqrt{c_2^2+c_1^2B^2}}, \nonumber \\
                                                                                \\
\mathbf{e}_2'=\frac{-c_1B\mathbf{e}_1+c_2\mathbf{e}_2}{\sqrt{c_2^2+c_1^2B^2}}, \nonumber
\end{eqnarray}
where $B=1+2\sqrt{1+c_2^2/c_1^2}cos[(1/3)arccos(1/\sqrt{1+c_2^2/c_1^2})]$ is a real solution of the cubic equation $(c_1/c_2)^2B^3-3(c_1/c_2)^2B^2-3B+1=0$. In the new basis the Hamiltonian and dispersion are:
\begin{equation}
\label{ha3dr}
\widehat{H}=\left(E_0+c\mathbf{q}'^2\right)\widehat{\sigma}_0-\frac{c_1^2(B^2-2B)-c_2^2}{\sqrt{c_2^2+c_1^2B^2}}q_2'(q_2'^2-3q_1'^2)\widehat{\sigma}_1,
\end{equation}
\begin{equation}
\label{dis3dr}
E_{1,2}=E_0+c\mathbf{q}'^2\pm\frac{c_1^2(B^2-2B)-c_2^2}{\sqrt{c_2^2+c_1^2B^2}}|q_2'(q_2'^2-3q_1'^2)|.
\end{equation}
\endnumparts
Here we note that $c_1^2(B^2-2B)-c_2^2>0$. Since $q_2'(q_2'^2-3q_1'^2)=0$ has three solutions, three MNL intersects at $\mathbf{q}=0$ at angles of $\pi/3$.

Groups 33, 43, 45, $29^D$ and $33^D$ host fortune teller (FT) dispersion which is described in more detail in \cite{Ja17, Mi20}. Here we only give the dispersion:
\begin{equation}
E_{j,l}=j|c_1|q_1|+lc_2|q_2||, (j,l\in\left\{+,-\right\}),
\end{equation}
where $c_1$ and $c_2$ are (only in this formula) positive quantities. Bands are ordered in the following way: $E_{-,+}\leq E_{-,-}\leq E_{+,-}\leq E_{+,+}$. It follows that MNLs $E_{-,-}=E_{+,-}$ are given by $c_1q_1=c_2q_2$ and $c_1q_1=-c_2q_2$. Two MNLs intersect at $\mathbf{q}=0$ at an angle that is dependent on $c_1$ and $c_2$. Those MNLs arising from FT dispersion are at constant energy only in linear approximation. When quadratic corrections to the Hamiltonian are accounted for, the MNL remain unsplit but overall energy shift is quadratic in $q_{||}$ - the wave vector along MNL.

\section{Illustrative examples}

\begin{figure}
\includegraphics[width=\textwidth]{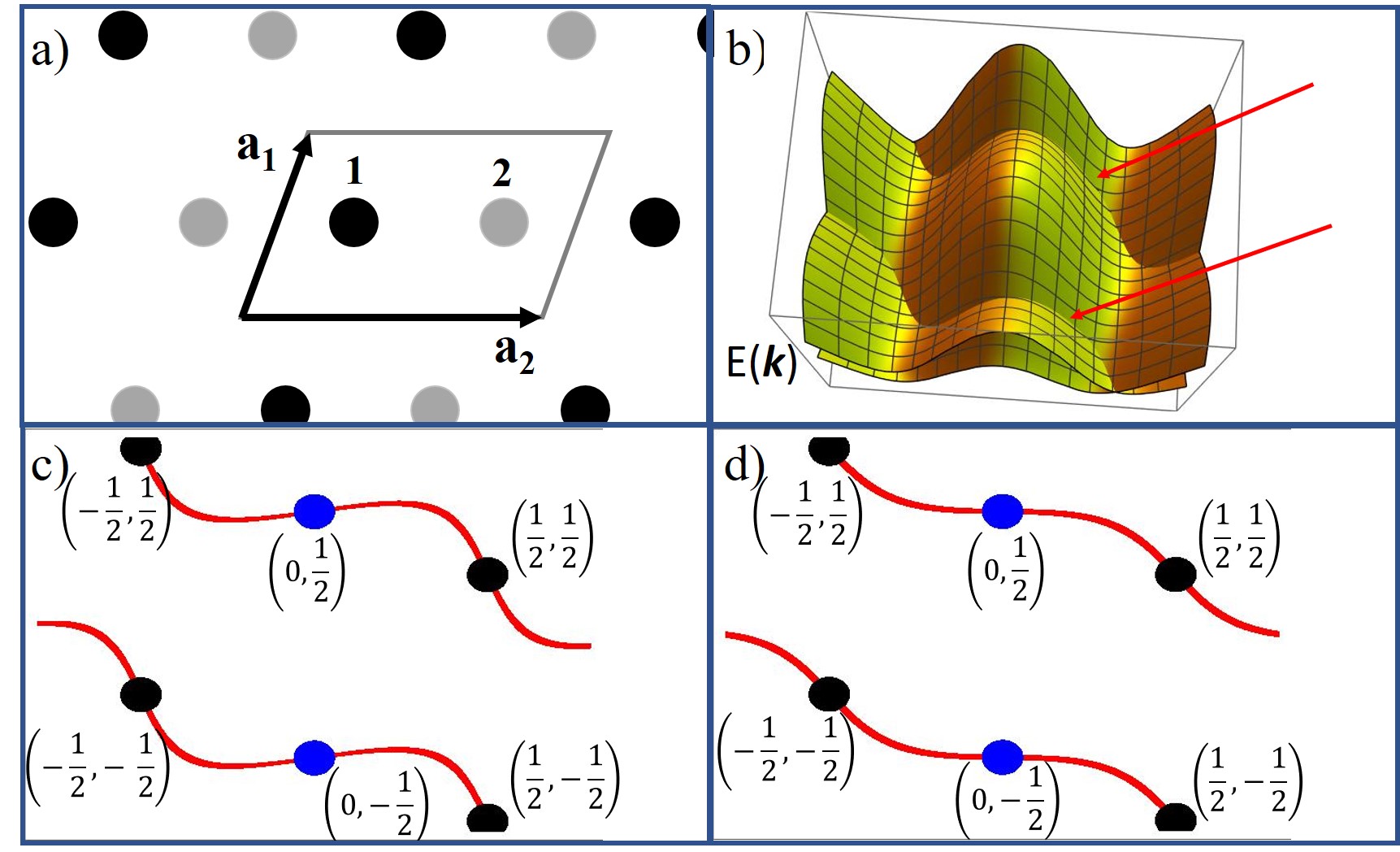}%
\caption{\label{Figu1} Tight binding model for $p112/b$: a) - crystal structure (all nuclei are of the same type, black and gray circles denotes nuclei above and below the drawing plane, respectively, visualization by VESTA \cite{vesta}), b) - electronic band structure for $f_0$, $f_1$, $f_2$ and $f_3$ equal to $4.2$, $1.3$, $0.9$ and $0.5$ in arbitrary units ($a.u.$), respectively (red arrows denote MNLs), c) - position of MNL in the reciprocal space, d) - same as c) for $f_0$, $f_1$, $f_2$ and $f_3$ equal to $3.6$, $3.1$, $2.7$ and $0.9$ $a.u.$, respectively. In c) and d) numbers in brackets denote coordinates of special points in the reciprocal basis (blue and black circles). Mutually equivalent points are denoted with circles of the same color.}
\end{figure}

We first consider the tight-binding model from the $s$-orbitals, without SOC, on the structure shown in Figure \ref{Figu1}a), that belongs to layer group $p112/b$ ($7$). Occupied Wyckoff position is $2d$ with $z=1.3$\AA. The Hamiltonian and its eigenvalues are:
\numparts
\label{tbm7}
\small
\begin{eqnarray}
\label{tbm7H}
\fl\widehat{H}(\mathbf{k})= \\
\fl\left(
\begin{array}[c]{cc}
	f_0+2f_1cos(\mathbf{k}\cdot\mathbf{a}_1) & 2e^{-\frac{i}{2}\mathbf{k}\cdot\mathbf{a}_2}(f_2cos(\frac{1}{2}\mathbf{k}\cdot\mathbf{a}_2)+f_3cos(\mathbf{k}\cdot\mathbf{a}_1-\frac{1}{2}\mathbf{k}\cdot\mathbf{a}_2)) \\
	2e^{\frac{i}{2}\mathbf{k}\cdot\mathbf{a}_2}(f_2cos(\frac{1}{2}\mathbf{k}\cdot\mathbf{a}_2)+f_3cos(\mathbf{k}\cdot\mathbf{a}_1-\frac{1}{2}\mathbf{k}\cdot\mathbf{a}_2)) & f_0+2f_1cos(\mathbf{k}\cdot\mathbf{a}_1)
\end{array}
\right), \nonumber
\end{eqnarray}
\normalsize
\begin{equation}
\label{tbm7D}
\fl E_{1,2}(\mathbf{k})=f_0+2f_1cos(\mathbf{k}\cdot\mathbf{a}_1)\pm2|f_2cos(\frac{1}{2}\mathbf{k}\cdot\mathbf{a}_2)+f_3cos(\mathbf{k}\cdot\mathbf{a}_1-\frac{1}{2}\mathbf{k}\cdot\mathbf{a}_2)|.
\end{equation}
Here $f_0$, $f_1$, $f_2$ and $f_3$ are hopping real parameters for zeroth, first, second and third neighbors, respectively. The two energies coincide for:
\begin{equation}
\label{tbm7uslov}
\mathbf{k}\cdot\mathbf{a}_2\in\left\{-2arctan\left(\frac{f_2+f_3cos(\mathbf{k}\cdot\mathbf{a}_1)}{f_3sin(\mathbf{k}\cdot\mathbf{a}_1)}\right)+2m\pi\right\},
\end{equation}
\endnumparts
where $m$ is any integer. Equation (\ref{tbm7uslov}) is well defined for all possible values of hopping parameters. The full band structure for particular values of parameters is shown in Figure \ref{Figu1}b), where band touching lines are clearly visible. Position of lines in the reciprocal space are shown in Figure \ref{Figu1}c). By varying parameters, the shape of lines changes but the lines always pass through points $(0,1/2)$ and $(1/2,1/2)$ and equivalent to them, as shown in Table \ref{tbl1} and Figure \ref{Figu1}c), d).

\begin{figure}
\includegraphics[width=\textwidth]{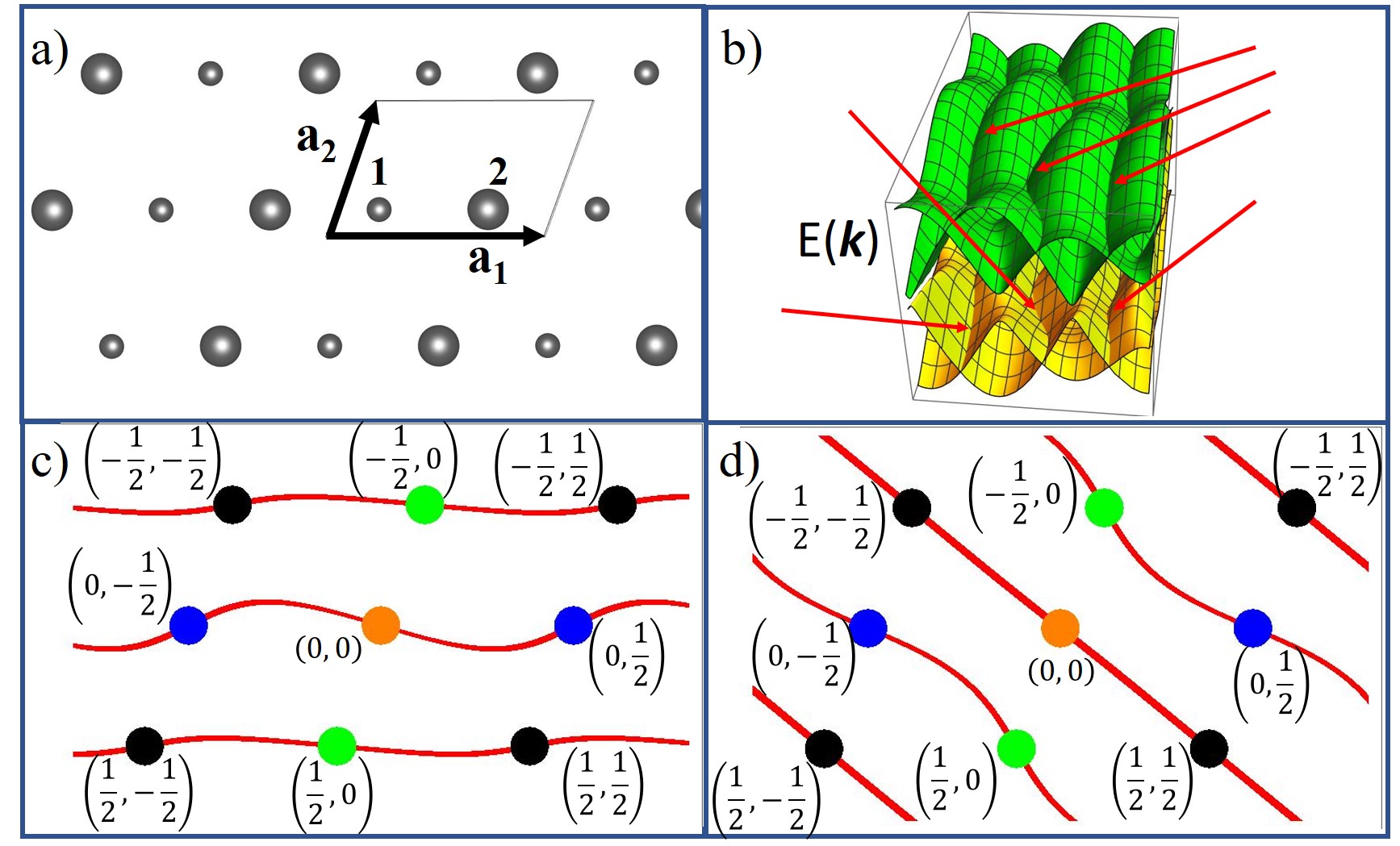}%
\caption{\label{Figu2} Tight binding model for symmorphic $p11m$: a) - crystal structure (nuclei of both types are located at one plane, visualization by VESTA \cite{vesta}), b) - electronic band structure for $f_0$, $f_1$, $z_1$, $z_2$ and $z_3$ equal to $2$, $3$, $4+2i$, $3-6i$ and $1+2i$ $a.u.$, respectively (red arrows denote MNLs), c) - position of MNL in the reciprocal space, d) - same as c) for $f_0$, $f_1$, $z_1$, $z_2$ and $z_3$ equal to $4.2$, $1.3$, $-2.4+1.2i$, $3.4-5.6i$ and $6.1+1.2i$ $a.u.$, respectively. In c) and d) numbers in brackets denote coordinates of special points in the reciprocal basis (blue, orange, green and black circles). Mutually equivalent points are denoted with circles of the same color.}
\end{figure}

We next consider the tight-binding model on the structure shown in Figure \ref{Figu2}a) that belongs to symmorpic layer (double) group $p11m$ ($4^D$). We choose basis functions $\left|s_1\uparrow\right\rangle, \left|s_1\downarrow\right\rangle, \left|s_2\uparrow\right\rangle, \left|s_2\downarrow\right\rangle$, where $s_1, s_2$ are $s$-orbitals for atoms $1$ and $2$, respectively, while $\uparrow$ and $\downarrow$ denote spinors for up and down spin, respectively. Since horizontal reflection plane is symmetry of the system, the \emph{up} and \emph{down} states are decoupled from each other, so that Hamiltonian is block-diagonal:
\numparts
\label{tbm4dou}
\begin{equation}
\widehat{H}=\left(
\begin{array}[c]{cc}
	\widehat{H}_{\uparrow} & \widehat{0} \\
	\widehat{0} & \widehat{H}_{\downarrow}
\end{array}
\right),
\end{equation}
with
\begin{equation}
\fl\widehat{H}_{\uparrow}(\mathbf{k})=\widehat{H}_{\downarrow}^*(-\mathbf{k})=\left(
\begin{array}[c]{cc}
	f_0 & z_1e^{-i\mathbf{k}\cdot\mathbf{a}_1}+z_2+z_3e^{i\mathbf{k}\cdot(\mathbf{a}_2-\mathbf{a_1})} \\
	z_1^*e^{i\mathbf{k}\cdot\mathbf{a}_1}+z_2^*+z_3^*e^{-i\mathbf{k}\cdot(\mathbf{a}_2-\mathbf{a_1})} & g_0
\end{array}
\right).
\end{equation}
Here $f_0$ and $g_0$ are real, on-site hopping parameters for sites $1$ and $2$, respectively, while $z_1$, $z_2$ and $z_3$ are complex hoping parameters for first, second and third neighbors, respectively. The full band structure is:
\begin{equation}
E_2^{\uparrow, \downarrow}(\mathbf{k})=\frac{f_0+g_0}{2}+\frac{1}{2}\sqrt{(f_0-g_0)^2+4|z_1e^{\mp i\mathbf{k}\cdot\mathbf{a}_1}+z_2+z_3e^{\pm i\mathbf{k}\cdot(\mathbf{a}_2-\mathbf{a}_1)}|^2},
\end{equation}
\begin{equation}
E_1^{\uparrow, \downarrow}(\mathbf{k})=\frac{f_0+g_0}{2}-\frac{1}{2}\sqrt{(f_0-g_0)^2+4|z_1e^{\mp i\mathbf{k}\cdot\mathbf{a}_1}+z_2+z_3e^{\pm i\mathbf{k}\cdot(\mathbf{a}_2-\mathbf{a}_1)}|^2}.
\end{equation}
Both nodal lines $E_1^{\uparrow}=E_1^{\downarrow}$ and $E_2^{\uparrow}=E_2^{\downarrow}$ are given by the equation:
\begin{equation}
\label{usl4dou}
h_3sin(\mathbf{k}\cdot\mathbf{a}_1)+h_2sin(\mathbf{k}\cdot\mathbf{a}_2)+h_1sin[\mathbf{k}\cdot(\mathbf{a}_2-\mathbf{a}_1)]=0,
\end{equation}
where $h_1=-i(z_2^*z_3-z_2z_3^*)$, $h_2=-i(z_1^*z_3-z_1z_3^*)$ and $h_3=-i(z_1^*z_2-z_1z_2^*)$ are real parameters. Equation (\ref{usl4dou}) has the following solution:
\begin{equation}
\label{resusl}
\fl\mathbf{k}\cdot\mathbf{a}_2\in\left\{\Phi-arcsin\left(\frac{h_3}{h_1}sin\Phi\right)+2m\pi,\Phi+arcsin\left(\frac{h_3}{h_1}sin\Phi\right)+(2m+1)\pi\right\},
\end{equation}
with $m$ being any integer and
\begin{equation}
sin(\Phi)=\frac{h_1sin(\mathbf{k}\cdot\mathbf{a}_1)/h_3}{\sqrt{(h_1^2+h_2^2)/h_3^2+2h_1h_2cos(\mathbf{k}\cdot\mathbf{a}_1)/h_3^2}},
\end{equation}
\begin{equation}
cos(\Phi)=\frac{h_1cos(\mathbf{k}\cdot\mathbf{a}_1)/h_3+h_2/h_3}{\sqrt{(h_1^2+h_2^2)/h_3^2+2h_1h_2cos(\mathbf{k}\cdot\mathbf{a}_1)/h_3^2}}.
\end{equation}
\endnumparts
The argument of $arcsin$ in (\ref{resusl}) has modulus less or equal to one except if $(h_1/h_3)^2$ and $(h_2/h_3)^2$ are both less then one. In that case $-1\leq cos(\mathbf{k}\cdot\mathbf{a}_1)\leq t_2$ or $t_1\leq cos(\mathbf{k}\cdot\mathbf{a}_1)\leq 1$ must hold ($t_{1,2}=-h_1h_2/h_3^2\pm\sqrt{(1-h_1^2/h_3^2)(1-h_2^2/h_3^2)}$). Band structure shown in Figure \ref{Figu2}b) for particular values of parameters, confirm existence of MNL. In this case $h_1^2$ and $h_2^2$ are both less then $h_3^2$, while in the case shown in Figure \ref{Figu2}d) this does not hold, with the clear difference in accidental degeneracy pattern. In any case, MNLs pass through HSPs as shown in Table \ref{tbl2} and in Figure \ref{Figu2}c), d).

\section{Discussion and conclusions}
We compare our results with the ones on related topics published in the literature. Analysis of all 3D crystallographic point double groups, for systems with SOC, that shows in which cases movable or unmovable nodal lines exist is reported \cite{kr1}. Their results can be used for symmorphic little groups, since then allowed representations are just a product of point group irreducible representations with the phase factors that represent translations. Point groups $\underline{C}_s^D$ and $\underline{C}_{3h}^D$ hosts one and three MNL, respectively \cite{kr1}. This is in accordance with our results in Table \ref{tbl2} (little group for $74^D$ in BZ center is $\underline{C}_{3h}^D$, for remaining symmorphic groups it is $\underline{C}_s^D$). On the other hand, HSPs for non-symmorphic $29^D$ and $33^D$ from Table \ref{tbl2} host two intersecting MNL, while \cite{kr1} predicts one nodal line along axis of order 2 for $\underline{C}_{2v}^D$. This difference is because $29^D$ and $33^D$ are non-symmorphic and because band degeneracy at HSPs hosting FT states is four, while \cite{kr1} considers two-band models.

Origin of doubly degenerate nodal lines that connect time-reversal invariant momenta in 3D crystals with SOC is reported in \cite{kr2}. List of symmorphic 3D space groups that host such lines \cite{kr2} contains all symmorphic corresponding space groups in Table \ref{tbl2}, which support our findings. Compound $\mathrm{PbTaSe_2}$ is a symmorphic 3D metal with SOC that exhibits nodal lines in experiments \cite{kr3}, as predicted in \cite{kr2}. To avoid Kramers degeneracy in the whole BZ, authors of \cite{kr2} considered non-centrosymmetric 3D crystals. They show that nodal line through HSP exists if it's little groups have at least one impropper rotation (symmetry element with determinant $-1$). In our cases this is always fulfilled, since at least horizontal plane must be present in order to avoid band repulsion due to non-crossing rule.

To illustrate their own results, authors of \cite{en23} used tight binding model with SOC on a structure that belongs to group $74^D$ ($p\overline{6}$). Three nodal lines intersects symmetrically at BZ-center and also pass through middle of BZ edges, thus confirming our prediction. Numerical calculations \cite{kr4} suggest that boron bilayer adopts structure with the symmetry $p\overline{6}$. Since boron is light element, reported band structure is without SOC \cite{kr4}. Future research could show if it is possible to enhance SOC using bilayer boron as a starting material. 

Group theoretical conditions for semi-Dirac dispersion in non-magnetic 2D materials with negligible SOC are reported \cite{JaSe17}.
All groups and HSPs from \cite{JaSe17} are here included in Table \ref{tbl1}. Tight-binding model for group $c\ 2/m\ 2/m\ 2/e$ gives line of degeneracy which depends on numerical values of tight-binding parameters \cite{JaSe17}, hence it is a MNL as predicted here. Moreover, the dispersion near $(0,1/2)$ point in \cite{JaSe17}, matches the one for group $c\ 2/m\ 2/m\ 2/e$ given here in (\ref{dis1dr}). Similarly, tight-binding model for group $p\ 1\ 1\ b$ published in \cite{JaSe18}, exhibits MNL across the whole BZ with the same dispersion near HSPs indicated in Table \ref{tbl1}. Lines of accidental degeneracy, noticed in \cite{JaSe18}, gained full explanation in the present work. Group $c\ m\ 2\ e$ is omitted in \cite{JaSe17}, due to a bug in REPRES program on Bilbao Crystallographic Server, which is fixed recently. According to \cite{JaSe17}, authors used REPRES \cite{BCS06} for finding allowed representations of layer groups via their corresponding space groups. Finally, one can point out that groups from \cite{JaSe17} give semi-Dirac dispersion (quadratic in one direction of the BZ that split linearly in the perpendicular direction) while the bands do not form cones. Formation of semi-Dirac cones requires Hamiltonian of the form: $c_1q_1\widehat{\sigma}_j+c_2q_2^2\widehat{\sigma}_l$ (or, in more general case: $c_1q_1\widehat{\sigma}_j+c_2q_2^n\widehat{\sigma}_l$), with two different Pauli matrices ($j\neq l$) and with $n$-natural number greater then one. We have omitted term with the unit matrix $\widehat{\sigma}_0$, since it does not lead to band splitting. In this case the band splitting is: $\pm \sqrt{c_1^2q_1^2+c_2^2q_2^4}$ (in more general case: $\pm \sqrt{c_1^2q_1^2+c_2^2q_2^{2n}}$), with $\mathbf{q}=0$ as the only touching point. Detail analysis of all HSP of all gray layer single or double groups does not give any HSP with semi-Dirac cones \cite{Mi22}.

In summary we have determined all layer groups that host nodal lines through high-symmetry BZ points, which shape depends on the model parameters (MNL). In our analysis we have included both cases without and with SOC in the presence of TRS. Although we didn't restrict our search to Weyl or Dirac nodal lines only, it turned out that all MNL we found are either DNL (in cases without SOC) or WNL (in cases with SOC). Since little group representations responsible for MNLs are the only allowed in listed HSPs, the existence of MNLs is guaranteed by the layer groups from Tables \ref{tbl1} and \ref{tbl2}. On the other hand, it is not guaranteed that MNLs are at constant energy, nor that they lie close to the Fermi level, although such cases are not necessarily excluded. Our finding extend the knowledge of band degeneracies in 2D materials and might be useful for designing 2D materials which sensitivity to external conditions like temperature, pressure, strain \emph{etc} is yet to be studied.

\ack{Author acknowledge funding by the Ministry of Science, Technological Development and Innovation of the Republic of Serbia provided by the Institute of Physics Belgrade, University of Belgrade.}

\section*{References}

\bibliographystyle{unsrt}

\end{document}